\documentclass[aps, prl, superscriptaddress, showkeys, showpacs, twocolumn, longbibliography, nofootinbib]{revtex4-2}  
\usepackage[utf8]{inputenc}
\usepackage{amsmath}
\usepackage{amsfonts}
\usepackage{amsthm}
\usepackage{amssymb}
\usepackage{float}
\usepackage{braket}
\usepackage{graphicx}
\usepackage{url}
\usepackage[colorlinks=true, pdfstartview=FitV, linkcolor=red, citecolor=blue, urlcolor=blue]{hyperref}
\usepackage{slashed}
\usepackage[normalem]{ulem}

\begin{document}

\title{Attractor of hydrodynamic attractors}

\author{Shile Chen}
\email[]{shile_chen@163.com}
\affiliation{Department of Physics, Tsinghua University, Beijing 100084, China.}

\author{Shuzhe Shi}
\email[]{shuzhe-shi@tsinghua.edu.cn}
\affiliation{Department of Physics, Tsinghua University, Beijing 100084, China.}
\affiliation{State Key Laboratory of Low-Dimensional Quantum Physics, Tsinghua University, Beijing 100084, China.}

\begin{abstract}
Understanding how hydrodynamics emerges rapidly in the medium produced by relativistic heavy-ion collisions remains a key theoretical challenge. While the attractor solution---manifesting as a non-thermal fixed point during the early evolution stage---offers a potential explanation, it does not fully account for how far-from-equilibrium systems quickly approach near-equilibrium states. In this Letter, we demonstrate that the attractor in a higher-order hydrodynamic theory converges to the same solution as a second-order theory before reaching the Navier--Stokes limit. This finding suggests that commonly used second-order hydrodynamic equations, provided they incorporate the correct transport coefficients, are adequate to approximate the system's behavior starting from an intermediate time---even when using a higher-order theory that would be more suitable for describing the far-from-equilibrium evolution.
\end{abstract}
\maketitle

\vspace{3mm}
\emph{Introduction}.--- 
Relativistic heavy-ion collision experiments provide a powerful tool for systematically studying the color-deconfined phase of matter---the quark-gluon plasma (QGP). A major achievement in heavy-ion phenomenology has been the remarkable success of numerical hydrodynamic simulations in describing and predicting measurements of collective hadronic behavior (see e.g., Refs.~\cite{Shuryak:2014zxa, Shen:2014vra, Schenke:2010rr, Karpenko:2013wva, vanderSchee:2013pia, Pang:2018zzo, Du:2019obx}). 

Relativistic hydrodynamic theory is built upon energy-momentum conservation, $\mathcal{D}_\mu T^{\mu\nu}=0$, with the stress tensor decomposed as\footnote{We work in the Landau frame where the fluid velocity is defined to be comoving with the energy, i.e., $T^{\mu\nu} u_\nu = \varepsilon\,u^\mu$.}
\begin{align}
    T^{\mu\nu} = (\varepsilon + P + \Pi) u^\mu u^\nu - (P+\Pi) g^{\mu\nu} + \pi^{\mu\nu}\,,
    \label{eq:hydro}
\end{align}
where the energy density ($\varepsilon$) and pressure ($P$) are related through the equation of state (EoS), $u^\mu$ is the flow velocity, and $\Pi$ (bulk pressure) and $\pi^{\mu\nu}$ (shear tensor) represent viscous corrections. Comparison between experimental data and hydrodynamic modeling has successfully validated the EoS from first-principles lattice QCD calculations~\cite{HotQCD:2014kol, Borsanyi:2010bp} and established a framework for investigating other QGP signatures like strangeness enhancement, quarkonium suppression, and jet quenching.

In commonly used hydrodynamic simulations~\cite{Shen:2014vra, Schenke:2010rr, Karpenko:2013wva, vanderSchee:2013pia, Pang:2018zzo, Du:2019obx}, the viscous terms in Eq.~\eqref{eq:hydro} evolve according to relaxation equations:
\begin{align}
    \tau_{\pi} u^\alpha \mathcal{D}_\alpha \pi^{\langle\mu\nu\rangle} + \pi^{\mu\nu}
    = 2\eta \sigma^{\mu\nu}
    - \delta_{\pi\pi}\theta \pi^{\mu\nu}
    - \tau_{\pi\pi} \pi^{\langle\mu \lambda}_{\;} \sigma_{\lambda}^{\nu\rangle}\,,
    \label{eq:hydro_vis}
\end{align}
where $\langle \cdot \cdot \rangle$ denotes the symmetrized, traceless spatial projection, $\theta\equiv \mathcal{D}_\mu u^\mu$ is the expansion rate, $\sigma^{\mu\nu} \equiv \mathcal{D}^{\langle\mu} u^{\nu\rangle}$ the shear gradient, $\eta$ the shear viscosity, and $\delta_{\pi\pi}$, $\tau_{\pi\pi}$ are second-order transport coefficients. Equation~\eqref{eq:hydro_vis} can be derived from both macroscopic~\cite{1959flme.book.....L, Baier:2007ix} and microscopic~\cite{Muller:1967zza, Israel:1979wp, Denicol:2012cn} approaches, which assume near-equilibrium conditions: the former employs thermodynamic principles with low-order gradient expansion, while the latter truncates the dynamic equations at linear order in non-equilibrium corrections of the phase-space distribution function. Consequently, hydrodynamics is traditionally considered valid only near local equilibrium, and its rapid applicability~\cite{Shen:2014vra, Nijs:2020roc, Nijs:2020ors} to heavy-ion collisions---which begin in a highly non-equilibrium state~\cite{Wang:1991hta, Gyulassy:1994ew, Schenke:2012wb}---remains theoretically puzzling.

In 2015, Heller and Spalinski~\cite{Heller:2015dha} proposed that attractor solutions might resolve this puzzle. Studying boost-invariant, transversely homogeneous systems, they showed that regardless of initial viscous corrections, all solutions rapidly converge to a universal attractor within the hydrodynamic regime. These non-thermal fixed points~\cite{Heller:2023mah} appear not only in heavy-ion collisions~\cite{Baier:2000sb, Berges:2013fga, Kurkela:2015qoa}, but across diverse systems spanning energy scales---from ultracold quantum gases~\cite{Prufer:2018hto, Erne:2018gmz, doi:10.1126/science.aat5793, Martirosyan:2023mml, Heller:2025yxm} to the early universe~\cite{Micha:2002ey, Berges:2008wm}. See~\cite{Heller:2013fn, Romatschke:2017vte, Romatschke:2017acs, Blaizot:2017ucy, Spalinski:2017mel, Ambrus:2021sjg, Kurkela:2019set, Almaalol:2020rnu, Blaizot:2019scw, Blaizot:2020gql, Blaizot:2021cdv, Chen:2024pez} for more developments, as well as reviews in~\cite{Romatschke:2017ejr, Alqahtani:2017mhy, Florkowski:2017olj, Berges:2020fwq, Shen:2020mgh}. The hydrodynamic equations exhibit distinct fast (decaying) and slow modes, with initial conditions affecting only the transient fast modes while all solutions asymptotically follow the same slow mode---a behavior independent of initial conditions.

Although attractors provide valuable insight into the transition from far-from-equilibrium to near-equilibrium dynamics, existing formulations are based on second-order hydrodynamic theories that themselves assume near-equilibrium conditions. While a complete description of far-from-equilibrium evolution would require higher-order theories that incorporate full collision kernels and higher moments of the distribution function~\cite{Bazow:2015dha, Lu:2025yry}, it remains unclear whether such higher-order theories would converge to the same attractor solution.

In this Letter, we bridge this conceptual gap by demonstrating that properly formulated second-order hydrodynamics can effectively approximate higher-order theories from intermediate times onward. Unlike previous studies focusing on attractors within individual theories, we establish a correspondence between different orders of hydrodynamic formulations. Our conclusion stems from systematically comparing attractors across hydrodynamic orders and analyzing late-time asymptotics via slow-roll expansion~\cite{Heller:2015dha, Denicol:2017lxn}.

\vspace{3mm}
\emph{Higher-order non-linear equations}.---
Following previous approaches, we consider conformal systems with longitudinal boost invariance and transverse homogeneity. For such systems, the thermodynamic relations simplify to $\varepsilon = 3P = \frac{3}{4}sT$, the expansion rate becomes $\theta = 1/\tau$, the bulk pressure vanishes, and the non-zero components of the shear velocity gradient satisfy $-2\sigma^{x}_{x} = -2\sigma^{y}_{y} = \sigma^{\eta}_{\eta} = \frac{2}{3\tau}$. We define $\varpi \equiv -2\pi^{x}_{x} = -2\pi^{y}_{y} = \pi^{\eta}_{\eta}$ as the only non-vanishing components of the shear viscous tensor.

The energy conservation equation takes the form:
\begin{align}
    \tau \partial_\tau \varepsilon = -\frac{4}{3} \varepsilon + \varpi\,,
    \label{eq:hydro_e}
\end{align}
where $\varpi = \varpi_{\mathrm{NS}} \equiv \frac{4C_\eta s}{3\tau}$ in the first-order, Navier--Stokes (NS) theory. In higher-order hydrodynamic theories, $\varpi$ evolves according to more complex equations of motion which are described as follows.

Several hydrodynamic frameworks---including second-order theories like Müller--Israel--Stewart (MIS)~\cite{Muller:1967zza, Israel:1979wp} and Denicol--Niemi--Moln\'ar--Rischke (DNMR)~\cite{Denicol:2012cn, Strickland:2017kux}, as well as third-order extensions like Diles--Mamani--Miranda--Zanchin (DMMZ)~\cite{Diles:2019uft} and Panday--Jaiswal--Patra (PJP)~\cite{Panday:2024hqp}---describe $\varpi$ through a standalone equation:
\begin{align}
\begin{split}
    &\tau_\pi \partial_\tau \varpi + \varpi - \frac{4\eta}{3\tau} + \frac{4}{3} \frac{\tau_\pi \varpi}{\tau}
\\=\;& 
     - c_{2,1} \frac{\tau_\pi \varpi}{\tau}
     + c_{2,2} \frac{\varpi^2}{\varepsilon}
     - c_{3,1} \frac{\tau^2_\pi\varpi}{\tau^2}
     - c_{3,2} \frac{\tau_\pi\varpi^2}{\tau\varepsilon}
     \,,
\end{split}\label{eq:hydro}
\end{align}
where $\tau_\pi$ is the shear stress relaxation time, and $c_{i,j}$ are theory-dependent transport coefficients (see Table~\ref{tab:coefficients_1}). For conformal systems, the $c_{2,2} \frac{\varpi^2}{\varepsilon}$ term captures all nonlinear effects from elastic scattering. While DNMR theory predicts $c_{2,2} = \frac{27}{70}$ for hard-sphere collisions~\cite{Denicol:2012cn, Denicol:2014vaa}, we adopt the relaxation-time approximation from Ref.~\cite{Strickland:2017kux} for better cross-theory comparison (see Supplemental Materials for $c_{2,2}$ effects).

\begin{table}[!hbpt]\centering
\begin{tabular}{l|c|c|c|c|c}
\hline\hline
    Model & $c_{2,1}$ & $c_{2,2}$ & $c_{3,1}$ & $c_{3,2}$ \\
\hline
MIS~\cite{Muller:1967zza, Israel:1979wp}
       & $0$ & $-\frac{3C_\lambda}{8C_\eta}$ & $0$ & $0$ \\
DNMR~\cite{Denicol:2012cn, Strickland:2017kux} 
       & $\frac{10}{21}$ & $0$ & $0$ & $0$\\
PJP~\cite{Panday:2024hqp}
       & $\frac{10}{21}$ & $0$ & $\frac{8}{63}$  & $\frac{89}{49}$\\
\hline\hline
\end{tabular}
\caption{Transport coefficients in second- and third-order hydrodynamic theories (Eqs.~\eqref{eq:hydro} and~\eqref{eq:attractor}). Note that Ref.~\cite{Diles:2019uft} does not specify coefficients.}
\label{tab:coefficients_1}
\end{table}

Following Heller and Spalinski~\cite{Heller:2015dha}, we introduce the anisotropy parameter $\varphi \equiv \varpi/\varepsilon$ and scaled proper time $w \equiv \tau T$. This transforms Eqs.~\eqref{eq:hydro_e} and~\eqref{eq:hydro} into the constraint equation $\tau\frac{\partial_\tau  \varepsilon}{\varepsilon} = \varphi - \frac{4}{3}$ and
\begin{align}
\begin{split}
    &\Big(\frac{2}{3}+\frac{\varphi}{4}\Big)\partial_w \varphi 
    + \frac{\varphi-\varphi_\mathrm{NS}}{C_\tau}
    + \frac{\varphi^2}{w}
\\=\;&
    - c_{2,1} \frac{\varphi}{w}
    + c_{2,2} \frac{\varphi^2}{C_\tau} 
    - c_{3,1} \frac{C_\tau\varphi}{w^2}
    - c_{3,2} \frac{\varphi^2}{w}\,,
\end{split}\label{eq:attractor}
\end{align}
where $\varphi_\mathrm{NS} = \frac{16C_\eta}{9w}$. Attractor solutions of these equations have been studied in Refs.~\cite{Heller:2015dha, Strickland:2017kux}.

\begin{figure}[!h]
    \centering
    \includegraphics[width=0.45\textwidth]{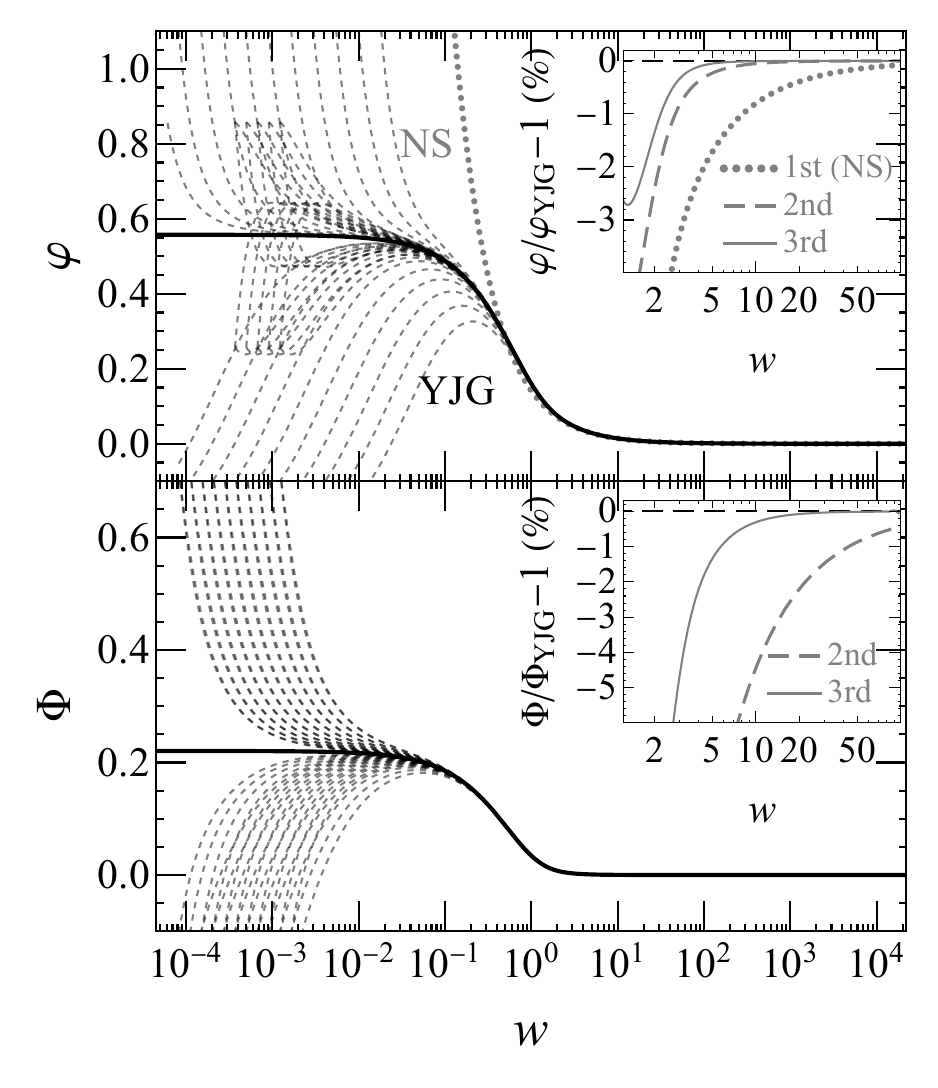}
    \caption{Attractor solutions in YJG theory [Eq.~\eqref{eq:attractor_higher}]. Top: anisotropy parameter $\varphi$; Bottom: higher-rank moment $\Phi$. Dashed curves show different initial conditions, solid curves the attractor, and dotted curves the NS solution. Insets compare first- (dotted), second- (dashed), and third-order (solid) expansions in Eq.~\eqref{eq:attractorExpansion_c}.}
    \label{fig:YJG}
\end{figure}

Recent developments by Brito--Denicol (BD)~\cite{deBrito:2023tgb} and Ye--Jeon--Gale (YJG)~\cite{Ye:2024phs} incorporate higher-order four-index tensors to better capture non-equilibrium effects. Different from the standalone equation~\eqref{eq:attractor}, they yield coupled equations for the attractors:
\begin{align}
\begin{split}
    &\Big(\frac{2}{3}+\frac{\varphi}{4}\Big)\partial_w \varphi 
    + \frac{\varphi-\varphi_\mathrm{NS}}{C_\tau}
    + \frac{\varphi^2}{w}
\\=\;&
    - c_{2,1}\frac{\varphi}{w}
    + c_{2,2}\frac{\varphi^2}{C_\tau} 
    + c_{\varphi,\Phi} \frac{\Phi}{w}
\,,\\~\\
    &\Big(\frac{2}{3}+\frac{\varphi}{4}\Big)\partial_w \Phi 
    + \Big(\frac{\Phi}{C_\Phi} - c_{\Phi,\varphi}\frac{\varphi}{w}\Big)
\\=\;& -c_{\Phi,\Phi}\frac{\Phi}{w} - c_{\Phi,2}\frac{\varphi\,\Phi}{w}\,,
\end{split} \label{eq:attractor_higher}
\end{align}
Here $\Phi$ represents the rank-four non-equilibrium moment , with relaxation time $\tau_\Phi = C_\Phi /T$. $\Phi$ is equivalent to $\Theta^{\eta\eta}_{\eta\eta}/(\varepsilon T^2)$ in BD and $\varsigma^{\eta\eta}_{\eta\eta}/\varepsilon$ in YJG, which are higher-order moments in the viscous correction $\Theta^{\eta\eta}_{\eta\eta} = \int \frac{(f-f_\mathrm{eq})\,d^3p}{(2\pi)^3 E_p} p^{\langle\eta}p^{\eta}p_\eta p_{\eta\rangle}$, and $\varsigma^{\eta\eta}_{\eta\eta} = \int \frac{(f-f_\mathrm{eq})\,d^3p}{(2\pi)^3 E_p (u\cdot p)^2} p^{\langle\eta}p^{\eta}p_\eta p_{\eta\rangle}$. Table~\ref{tab:coefficients_higher} lists the corresponding transport coefficients.

\begin{table}[!hbpt]\centering
\begin{tabular}{l|c|c|c|c|c|c}
\hline\hline
    Model & $c_{2,1}$ & $c_{2,2}$ & $c_{\varphi,\Phi}$ & $c_{\Phi,\varphi}$ & $c_{\Phi,\Phi}$ & $c_{\Phi,2}$\\
\hline
BD~\cite{deBrito:2023tgb}
       & $\frac{10}{21}$ & $0$ & $-\frac{1}{72}$ & $\frac{768}{35}$ & $\frac{60}{77}$ & $\frac{3}{2}$\\
YJG~\cite{Ye:2024phs} 
       & $\frac{10}{21}$ & $0$ & 1 & $\frac{96}{245}$ & $\frac{100}{231}$ & $1$  \\
\hline\hline
\end{tabular}
\caption{Transport coefficients in BD~\cite{deBrito:2023tgb} and YJG~\cite{Ye:2024phs} theories [Eq.~\eqref{eq:attractor_higher}].}
\label{tab:coefficients_higher}
\end{table}

Figure~\ref{fig:YJG} demonstrates the attractor behavior in YJG theory (using $C_\eta = 1/4\pi$, $C_\tau = C_\Phi = 5C_\eta$), where all solutions converge to the attractor---defined according to the asymptotic $w\partial_w\varphi = w\partial_w\Phi = 0$ at $w\to0$---before approaching the NS limit, regardless of initial conditions. 

\vspace{3mm}
\emph{Attractors and asymptotic behaviors}.---
The convergence of $\varphi-w$ trajectories from different initial conditions can be understood by analyzing the late-time asymptotic behavior of the viscous hydrodynamic equations. Following Ref.~\cite{Heller:2015dha}, we perform a slow-roll expansion in powers of $1/w$ for the anisotropy parameter:
\begin{align}
\begin{split}
\varphi = \sum_{n=0}^3\frac{A^\mathrm{s}_{n}}{w^n}  +\mathcal{O}\big(w^{-4}\big),
\end{split}\label{eq:attractorExpansion_s}
\end{align}

Substituting this into Eq.~\eqref{eq:attractor} yields the expansion coefficients\footnote{An alternative solution $A_{0} = 1/c_{2,2}$ exists when $c_{2,2} \neq 0$, but corresponds to an unstable fixed point.}:
\begin{align}
\begin{split}
A^\mathrm{s}_{0} &=\; 0,\\
A^\mathrm{s}_{1} &=\; \frac{16C_\eta}{9}\,,\\
A^\mathrm{s}_{2} &=\; \frac{16C_\eta}{9}\Big((\frac{2}{3}-c_{2,1}) C_\tau + \frac{16}{9} c_{2,2} C_\eta\Big),\\
A^\mathrm{s}_{3} &=\; \frac{16C_\eta}{9}\Big[
    (\frac{8}{9} - 2 c_{2,1} + c_{2,1}^2) C_\tau^2
    + \frac{512}{81} c_{2,2}^2 C_\eta^2 \\
&\quad + (- \frac{4}{3} + \frac{128}{27} c_{2,2} - \frac{16}{3} c_{2,1}c_{2,2} ) C_\eta C_\tau \\
&\quad - c_{3,1}  C_\tau^2 -\frac{16}{9} c_{3,2}  C_\eta C_\tau \Big].
\end{split}
\end{align}

For the coupled system (Eq.~\eqref{eq:attractor_higher}), we expand both $\varphi$ and $\Phi$:
\begin{align}
\begin{split}
\varphi &= \sum_{n=0}^3\frac{A_{n}^\mathrm{c}}{w^n}  +\mathcal O\big(w^{-4}\big),\\
\Phi &= \sum_{n=0}^3\frac{B_{n}^\mathrm{c}}{w^n}  +\mathcal O\big(w^{-4}\big),
\end{split}\label{eq:attractorExpansion_c}
\end{align}
yielding coefficients:
\begin{align}
\begin{split}
A_{0}^\mathrm{c} &=\; 0,\quad A_{1}^\mathrm{c} = \frac{16C_\eta}{9},\\
A_{2}^\mathrm{c} &=\; \frac{16C_\eta}{9}\Big((\frac{2}{3}-c_{2,1}) C_\tau + \frac{16}{9} c_{2,2} C_\eta\Big),\\
A_{3}^\mathrm{c} &=\; \frac{16\,C_\eta }{9}\Big[
     (\frac{8}{9} - 2 c_{2,1} + c_{2,1}^2)C_\tau^2
    + \frac{512}{81} c_{2,2}^2 C_\eta^2 \\
&\quad + (-\frac{4}{3} + \frac{128}{27} c_{2,2} - \frac{16}{3} c_{2,1}c_{2,2}) C_\eta C_\tau \\
&\quad + C_\Theta C_\tau c_{\varphi,\Phi} c_{\Phi,\varphi}\Big],
\end{split}
\end{align}
\begin{align}
\begin{split}
B_{0}^\mathrm{c} &=\; 0,\quad B_{1}^\mathrm{c} = 0,\\
B_{2}^\mathrm{c} &=\; \frac{16}{9}c_{\Phi,\varphi}\,C_\eta\,C_\Phi,\\
B_{3}^\mathrm{c} &=\; \frac{16}{9}c_{\Phi,\varphi}\,C_\eta\,C_\Phi \times \\
&\quad \Big(\frac{16}{9} c_{2,2} \, C_\eta + (\frac{4}{3} - c_{\Phi,\Phi}) C_\Phi + (\frac{2}{3}-c_{2,1}) C_\tau\Big),
\end{split}
\end{align}
where superscripts ``s" (standalone) and ``c" (coupled) distinguish the two cases.

Treating $C_\eta \sim C_\tau$ as a small expansion parameter, we find $A_n$ is proportional to $(C_\eta)^n$, ensuring the validity of the expansion. 
At order $1/w$, the solution is exactly the NS solution. Therefore, for all theories, their solutions and attractors converge to the NS limit at late times, regardless of the choice of $c_{i,j}$ coefficients. In the time interval before converging with $\varphi_\mathrm{NS}$, the ${A_2}/{w^{2}}$ term in Eqs.~\eqref{eq:attractorExpansion_s} and~\eqref{eq:attractorExpansion_c} dominates. 

The third-order transport coefficients ($c_{3,i}$, $c_{*,\Phi}$, and $c_{\Phi,*}$) appear only in $A_{n\geq3}$. Furthermore, $A_{3}^\mathrm{s}$ and $A_{3}^\mathrm{c}$ share the same dependence on second-order transport coefficients, differing only in terms involving third-order coefficients, indicating their similarity over a broader time range.

In Fig.~\ref{fig:YJG} subfigures, we compare the full solution with truncated expansions of Eq.~\eqref{eq:attractorExpansion_c} at first, second, and third order in $1/w$. Clearly, including higher-order terms improves agreement at smaller $w$ values.

While we focus on the YJG theory, attractor behaviors in second-order hydrodynamic theories and the BD theory were discussed in Refs.~\cite{Strickland:2017kux} and~\cite{deBrito:2023tgb}, respectively, and are re-examined in the Supplemental Material. Fig.~\ref{fig:attractors} (top) shows attractor solutions for all listed theories, with detailed comparisons in Fig.~\ref{fig:attractors} (bottom). Since DNMR, PJP, BD, and YJG theories share identical $c_{2,1}$ and $c_{2,2}$ values, they yield identical $A_2$ and similar $A_3$ coefficients in the slow-roll expansion. Although the MIS theory has different second-order parameters, it produces a similar $A_2$ (with $C_\lambda = 1/\pi$). Thus, we observe that: DNMR, PJP, BD, and YJG attractors converge by $w \gtrsim 20$; the MIS theory converges with them by $w \sim 100$; all approach the Navier--Stokes limit by $w\sim 400$.

\begin{figure}[!h]
    \centering
    \includegraphics[width=0.45\textwidth]{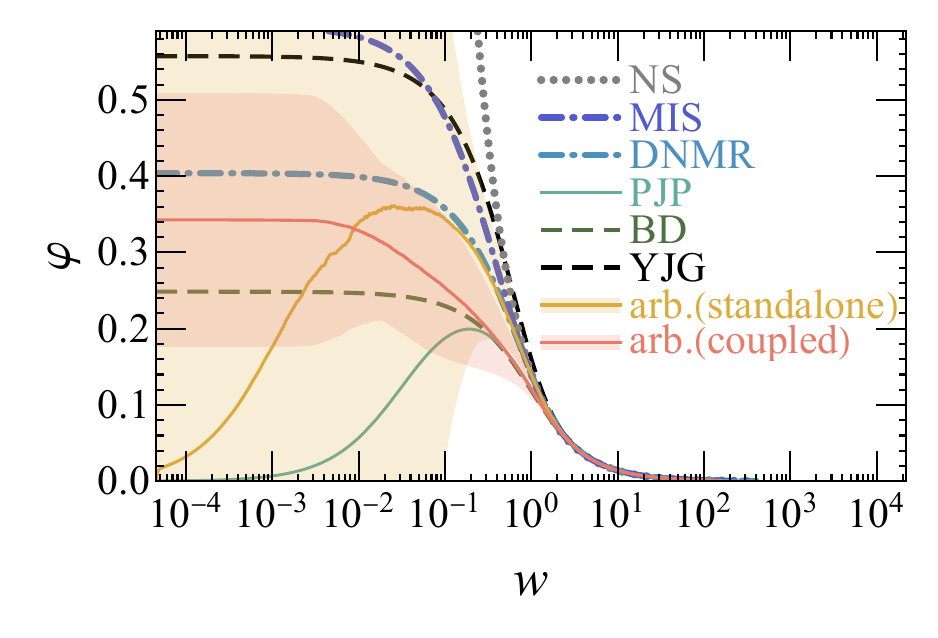}\\
    \includegraphics[width=0.45\textwidth]{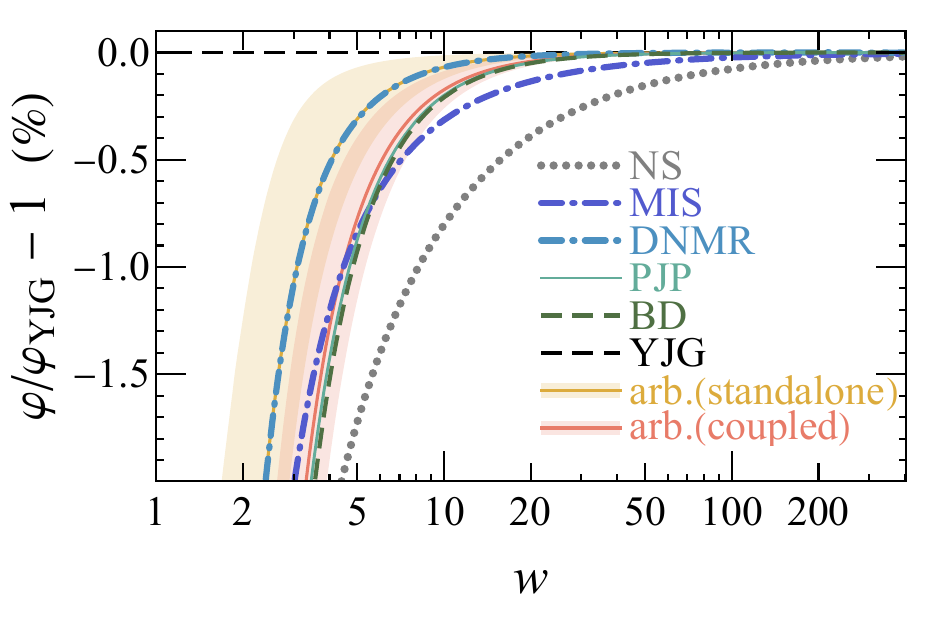}
    \caption{Attractor solutions for different theories (top) and their ratios relative to the YJG theory (bottom). The Navier--Stokes limit is shown as gray dotted lines for comparison.}
    \label{fig:attractors}
\end{figure}
The second-order slow-roll coefficients are identical in both equations ($A_2^\mathrm{s} = A_2^\mathrm{c}$), depending only on the second-order transport parameters $c_{2,1}$ and $c_{2,2}$ (besides $C_\eta$ and $C_\tau$). Consequently, theories sharing the same second-order transport coefficients---more precisely, the same combination $(c_{2,1} -\frac{16C_\eta}{9C_\tau} c_{2,2})$---exhibit identical late-time behavior before merging with $\varphi_\mathrm{NS}$.

To study theories with different microscopic interactions or non-equilibrium effects, we solve the attractors for both standalone~\eqref{eq:attractor} and coupled~\eqref{eq:attractor_higher} equations with fixed second-order coefficients ($c_{2,1}=10/21$, $c_{2,2}=0$) while randomly sampling higher-order coefficients:
\begin{align*}
c_{3,1} &\sim \mathcal{U}(-\tfrac{16}{63},\tfrac{16}{63}), & c_{3,2} &\sim \mathcal{U}(-\tfrac{178}{49},\tfrac{178}{49}), \\
c_{\varphi,\Phi} &\sim \mathcal{U}(0,2/27), & c_{\Phi,\varphi} &\sim \mathcal{U}(0,\tfrac{1536}{35}), \\
c_{\Phi,\Phi} &\sim \mathcal{U}(0,\tfrac{120}{77}), & c_{\Phi,2} &\sim \mathcal{U}(-3,3),
\end{align*}
where $\mathcal{U}(a,b)$ denotes a uniform distribution on $[a,b]$. These ranges match typical values in existing theories. For each case, we take $10^4$ samples and compute mean values and standard deviations at each $w$, shown as gold (standalone) and red (coupled) bands in Fig.~\ref{fig:attractors}. 

Despite starting from widely distributed initial conditions (evident from the broad initial bands), all theories converge to the same curve by $w\sim20$, matching the DNMR/PJP/BD/YJG convergence point. Before reaching the Navier--Stokes solution, the intermediate-time behavior depends primarily on $c_{2,1}$ and $c_{2,2}$, remaining insensitive to higher-order coefficients or equation structure.

We conclude that while far-from-equilibrium systems require higher-order hydrodynamics to fully capture momentum distribution details and nonlinear collision kernels, their behavior can be effectively approximated by second-order hydrodynamic theories when the second-order coefficients are properly chosen.

\vspace{3mm}
\emph{Summary and discussions}.---
Since their discovery in 2015, attractor solutions of hydrodynamic equations have emerged as a promising explanation for the rapid hydrodynamization of quark-gluon plasma in relativistic heavy-ion collisions. These solutions demonstrate how viscous corrections, regardless of their initial values, quickly converge to universal behaviors within the hydrodynamic regime. However, since conventional hydrodynamic equations are derived by truncating nonlinear non-equilibrium terms, the existence of attractors alone does not fully justify their applicability. 

In this Letter, we have advanced the understanding of this puzzle by demonstrating that higher-order hydrodynamic theories---which incorporate nonlinear viscous effects and additional degrees of freedom in the distribution function---exhibit identical late-time behavior to lower-order theories when they share the same second-order transport coefficients.

Focusing on boost-invariant longitudinally and transversely homogeneous systems, we have systematically studied attractor behavior across hydrodynamic orders. Both numerical solutions and slow-roll analysis of asymptotic behavior reveal that all theories with identical second-order coefficients converge to a universal attractor well before reaching the Navier--Stokes limit. This represents an ``attractor of attractors" across different theoretical formulations.

Our findings imply that for studying late-time evolution of the energy-momentum tensor---the relevant regime for heavy-ion collisions---second-order hydrodynamics with proper coefficients suffices to approximate the complete dynamics. This significantly simplifies practical applications while maintaining physical accuracy.

A natural next step for heavy-ion phenomenology involves relaxing spatial symmetry assumptions to investigate longitudinal and transverse structures, as initiated in~\cite{Romatschke:2017acs, Chen:2024pez, Chen:2024grb}. 

From the perspective of quantum thermalization in closed strongly-interacting systems, the insensitivity to fast-mode details underscores the dominant role of collective motion over microscopic quantum interactions. How initial quantum fluctuations give way to classical collective behavior remains a fundamental open question in our field.

\vspace{3mm}
\textbf{Acknowledgments.}--- We thank Lipei Du for helpful discussions. This work is supported by Tsinghua University under grant Nos. 04200500123, 531205006, 533305009. S.S. also acknowledge support by National Key Research and Development Program of China under Contract No. 2024YFA1610700.

\bibliography{ref}

\clearpage
\newpage 
\onecolumngrid 
\begin{appendix}
\section{Supplementary Material}
\twocolumngrid 
\section{More about Attractors}
\label{appendix1}

\begin{figure}[!h]
    \centering
    \includegraphics[width=0.45\textwidth]{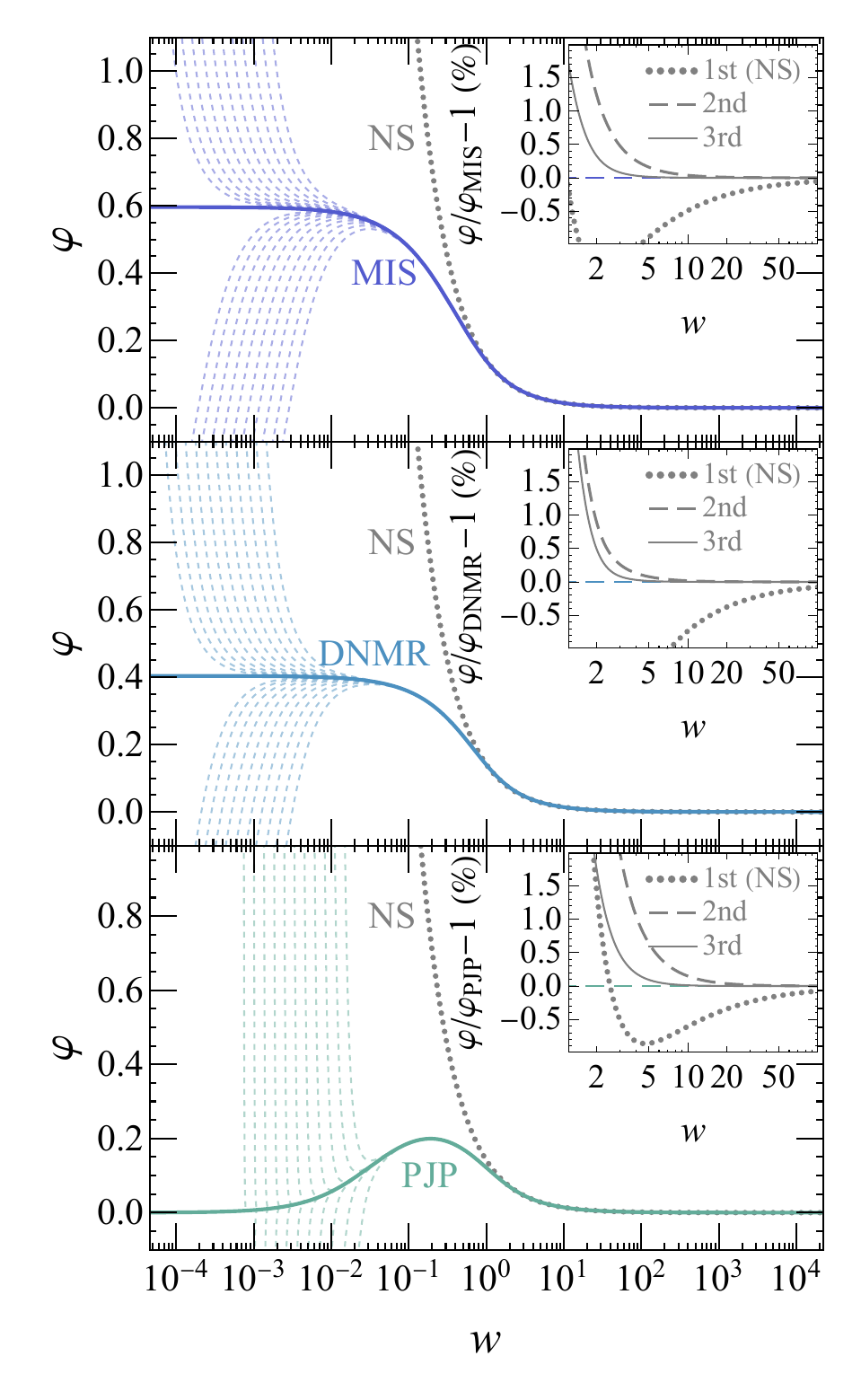}
    \caption{Attractor solutions for the standalone evolution equation of anisotropy~\eqref{eq:attractor}, showing convergence from different initial conditions (cf. Fig.~\ref{fig:YJG}).}
    \label{fig:app:standalone}
\end{figure}

\begin{figure}[!h]
    \centering
    \includegraphics[width=0.45\textwidth]{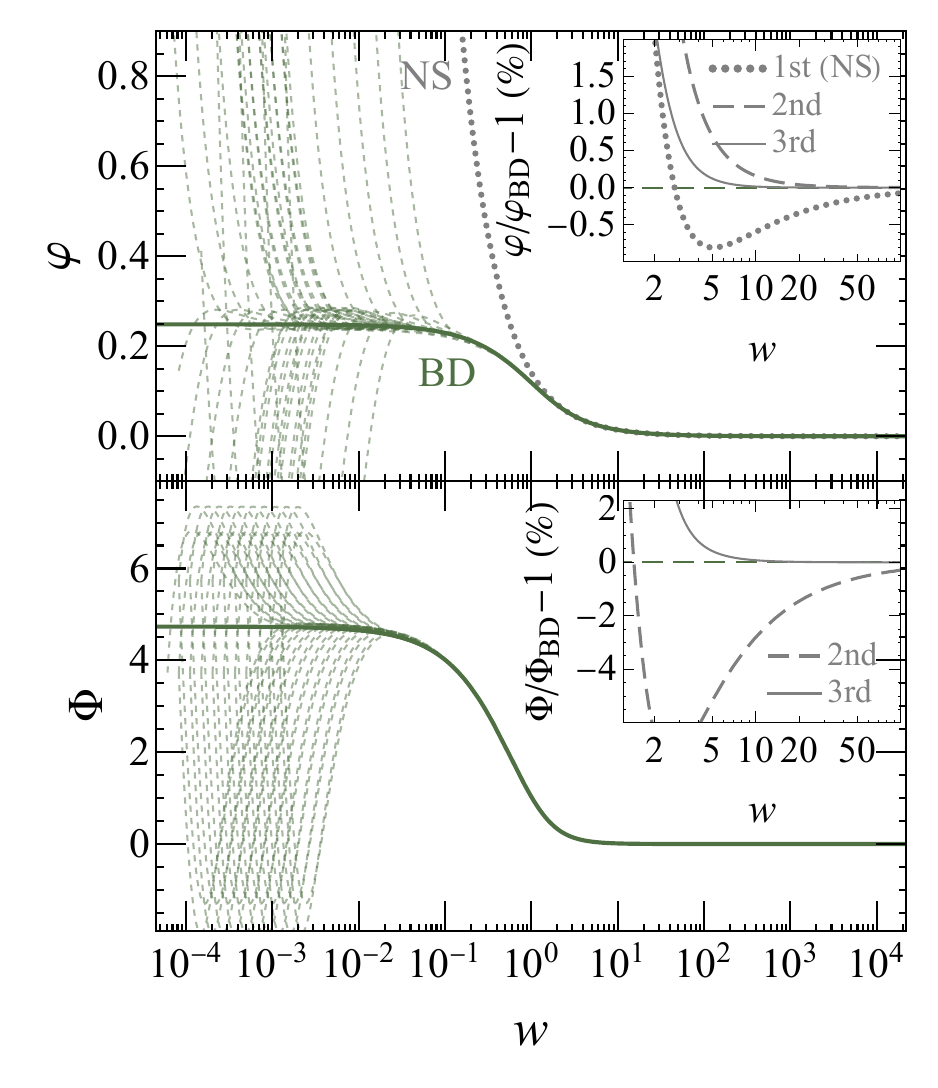}
    \caption{Attractor solutions for the BD theory (cf. Fig.~\ref{fig:YJG}).}
    \label{fig:app:BD}
\end{figure}

In this appendix, we examine attractor behavior across various hydrodynamic theories: MIS~\cite{Muller:1967zza, Israel:1979wp}, DNMR~\cite{Denicol:2012cn, Strickland:2017kux}, PJP~\cite{Panday:2024hqp}, and BD~\cite{deBrito:2023tgb}. Figures~\ref{fig:app:standalone} and~\ref{fig:app:BD} demonstrate that all theories exhibit convergence to their respective attractor solutions before approaching the Navier-Stokes limit. The slow-roll expansion shows excellent agreement with late-time evolution, with third-order expansions achieving $\sim 10^{-3}$ relative accuracy for $w \gtrsim 5$, compared to first-order Navier-Stokes solutions which only converge for $w \gtrsim 100$.

\begin{figure}[!h]
    \centering    
    \includegraphics[width=0.45\textwidth]{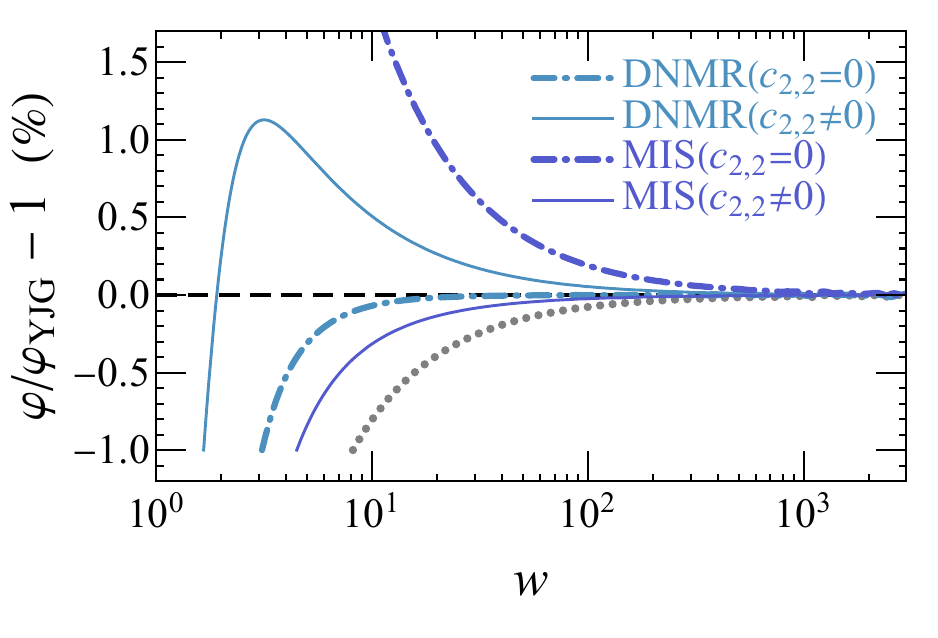}
    \caption{Comparison of DNMR (blue) and MIS (purple) theories with vanishing (dash-dot) versus nonvanishing (solid) $c_{2,2}$ coefficients (cf. Fig.~\ref{fig:attractors} lower panel).}
    \label{fig:second}
\end{figure}

As noted in the main text, different treatments of the $c_{2,2}$ coefficient exist in the literature for DNMR and MIS theories. Our slow-roll analysis reveals this parameter significantly affects long-time behavior. Figure~\ref{fig:second} highlights how proper $c_{2,2}$ values influence convergence rates.

\section{Early Time Behavior}
\label{appendix2}

While the main text analyzed late-time behavior via slow-roll expansion, we now examine early-time asymptotics. Notably, the PJP third-order hydrodynamics attractor originates from $\varphi(w=0)=0$, unlike other theories.

For the standalone equation~\eqref{eq:attractor}, setting $\varphi_0 \equiv \varphi(w\to0)$ and $w\partial_w\varphi =0$ yields:
\begin{align}
    (1+c_{3,2})\varphi_0^2 + c_{2,1} \varphi_0 - \frac{16C_\eta}{9C_\tau} = - c_{3,1} \frac{C_\tau\varphi_0}{w}.
\end{align}
This gives:
\begin{itemize}
    \item When $c_{3,1}\neq0$: $\varphi_0 \to \frac{16C_\eta w}{9c_{3,1}C_\tau^2}$;
    \item Otherwise: $\varphi_0 = \frac{\sqrt{c_{2,1}^2+\frac{64C_\eta}{9C_\tau}(1+c_{3,2})}-c_{2,1}}{2(1+c_{3,2})}$.
\end{itemize}

For coupled equations~\eqref{eq:attractor_higher}, we find:
\begin{align}
    \Phi(w\to0) = \frac{c_{\Phi,\varphi}\varphi_0}{c_{\Phi,\Phi}+ c_{\Phi,2} \varphi_0},
\end{align}
where $\varphi_0$ solves:
\begin{align}
    \varphi_0^2 + c_{2,1} \varphi_0 - \frac{c_{\varphi,\Phi} c_{\Phi,\varphi} \varphi_0}{c_{\Phi,\Phi}+ c_{\Phi,2} \varphi_0} = \frac{16C_\eta}{9C_\tau}.
\end{align}
\end{appendix}
\end{document}